# Algorithmization, requirements analysis and architectural challenges of TraConDa

## Tiamiyu, Osuolale Abdulrahamon[*]

Department of secured communication systems, St. Petersburg state university of telecommunications, Russia

## Abstract

*Globally, there are so much information security threats on Internet that even when data is encrypted, there is no guarantee that copy would not be available to third-party, and eventually be decrypted. Thus, trusted routing mechanism that inhibits availability of (encrypted or not) data being transferred to third-party is proposed in this paper. Algorithmization, requirements analysis and architectural challenges for its development are presented.*

## Keywords

*Trusted routing, secure data transfer, TraConDa, data security, data confidentiality, data integrity, security mechanism, unauthorized access prevention.*

## 1. Introduction

As recent development had shown that transferring confidential data over global data telecommunication network (GDTN) is no longer all that secure considering the fact that in best-selling routers are being embedded secretly different kind of "agents" for manipulation of traffic passing through those routers without the knowledge of the owner. Mostly this agents (implants) are embedded for the purpose of exploitation, to get "ungettable" [1]. Like the kind of halluxwater, stuccomontana, genie, gourmettrough, souffletrough, jetplow, sierramontana, headwater and feedtrough which could be found in cisco, juniper or Huawei networking devices [1, 2]. Some of these implants allow for complete control of router using hidden channels for information transmission. They are persistent. They survive an upgrade or replacement of the operating system, even some survive physically replacing the router's compact flash card. They modify operating system of networking device and enable covert functions to be remotely executed within the networking device via an Internet connection [1, 2].

Owing to the fact that communicating confidential, sometimes very critical, data over Internet is unavoidable; and the usage of Internet, which relies on telecommunication networks (TN), is ever increasing which exposes hundreds of millions of people worldwide to threats and risks, as Internet is vulnerable to cyber-attacks, it is paramount to have a mechanism of routing data from sender to recipient through those routing devices (RD) that are trusted Author in [3] confirmed that the routing control security mechanism is suitable for the realization of traffic confidentiality. Thus, as there are suitable security mechanisms described in GOST ISO R 7498-2-99 — Security Architecture that can be applied appropriately to protect the information exchanged between the application processes in the circumstances for which protection of communication between systems is required, those security mechanisms were analyzed to have in-depth knowledge on how routing control as a security mechanism is suitable for the realization of confidentiality and traffic confidentiality. This allows to understand that one of these protection mechanisms is a routing control mechanism, by which routes or paths can be chosen dynamically, or by such prior distribution, which uses only physically secured subnet, repeaters or data links [4]. Consequently, this routing control mechanism requires confidence (trust) in the intermediate nodes and therefore must be supplemented by a general architectural mechanism of trust functionality, to obtain fundamentally a new mechanism — a trusted routing (TR) mechanism (Fig. 1) which is process of planning and organizing information flow on a calculated route through TN nodes, excluding the possibility of tampering with the information in any

[*]Author for correspondence

form while the information stream is passing through
those TN nodes [5].

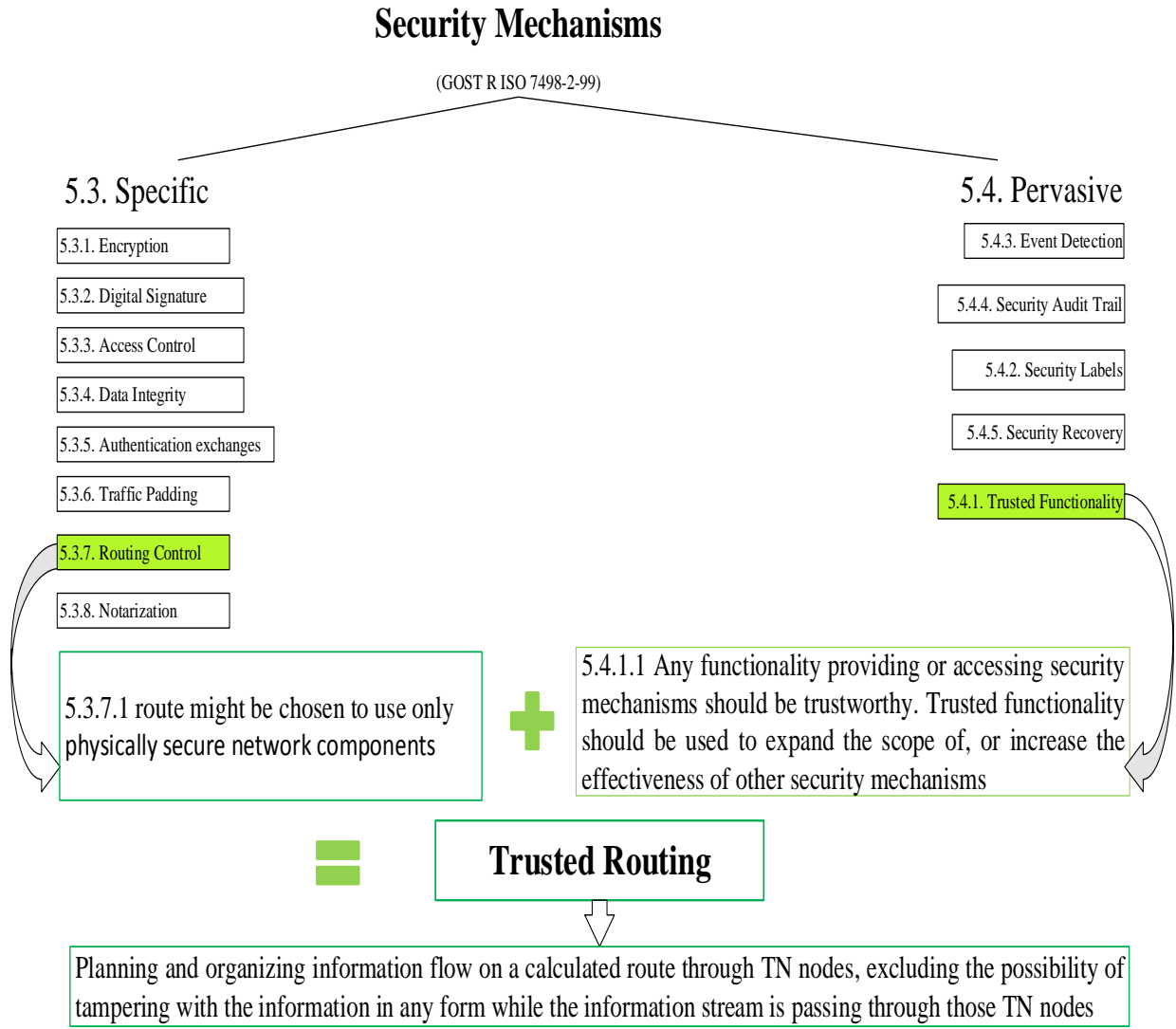

**Figure 1: Structure of Security Mechanisms revealing Trusted Routing Mechanism**

However, TR mechanism (that hereafter referred to as TraConDa — **Tra**nsfer **Con**fidential **Da**ta, for transferring confidential data securely in global data telecommunication network e.g. Internet) is not widely used in GDTN to combat information security threats (IST). One of the reasons is the lack of knowledge about its potentials, properties and implementation. Therefore it is necessary to deeply investigate the TR mechanism to provide evidence-based recommendations for its usage in GDTN and, if necessary, improvement. In view of the aforementioned, the purpose of this paper is establishing implementation methods, conditions, tools, and application boundaries for TR mechanism thereby presenting its architecture and requirements analysis. To achieve this goal, TR mechanism was investigated for its algorithmization, software implementation procedures, evaluation of its efficiency and recommendations for its improvement, which required addressing several research objectives, which are analyze the possibility of implementing TR mechanism in GDTN using "regular" tools in TCP/IP; if necessary, develop special tools for TR; synthesize software architecture management system for TR; evaluate the effectiveness of the TR mechanism in GDTN and make recommendations for its improvement. To achieve these research objectives are used modern research methods among which are collect, organize and analyze of scientific and technical information; systematic, cause-effective and comparative analysis; functional and structural synthesis; simulation; planning and processing of experimental results.

Since should routing control and trust functionality security mechanisms, described in GOST R ISO 7498-2-99 [6], are combined to have a kind of TR mechanism (Fig. 1), confidentiality and integrity of data being transferred on a network can be guaranteed. Further we analyze the possibility of implementing TR mechanism in GDTN using "regular" TCP/IP tools, and where necessary (when TCP/IP tool is not adequate), we develop special tool for TR; but before then, search for related works on TR mechanism is done to analyze other security mechanisms, for example MPLS VPN, that are having trust functionality.

### 1.1. Related trusted routing mechanism

Much work has been done to provide for data security via TR, among many known technics been studied is the one used in [7], method of secure routing based on degree of trust in which data can be routed through the chain of nodes, wherein at least one node has a lower confidence level, while the remaining nodes have acceptable levels of confidence. In the method is proposed creating and using an encrypted tunnel in the case when there is/are node(s) with an unacceptable level of trust and subsequently decoding the data in the TN when the node with an acceptable level of trust is found. However, the encrypted data, ultimately, can be copied or deleted by RD that does not inspire trust, and or be decrypted by third parties. There exist another method described in [8] in which the level of trust of nodes is determined when nodes establish trust by sending authentication messages encrypted with the shared secret key that was initially generated by a central authority and provisioned to nodes in a network for ensuing traffic encryption, and thereupon adding each other to their respective trust lists. In this method, in some cases, the problem arises when the exchange of general symmetric encryption key is compromised during attacks for the purpose of unauthorized access to information being transmitted. As a result, it is impossible to ensure the confidentiality and or integrity of data in TN during data transmission by providing the trusted path in TN through which the data passes from the sender to the recipient.

In [9], the authors propose a method of message security using trust-based multi-path routing. They give lower number of self-encrypted parts of a message to a fewer trusted nodes thereby making it difficult for malicious nodes to gain access to the minimum information required to break through the encryption strategy. Also they avoid unnecessary redundancy and non-trusted routes that may use brute force attacks and may decrypt messages if enough parts of the message are available to them, using trust levels.

For trusted computing, Jarrett, M. and Ward, P. [10] propose using trusted computing to prevent misconfigured or malicious nodes from participating in the network by providing additional security in open computing environments that allows software to prove its identity and integrity to remote entities. Closest to the technical solution by this proposed software model is [11, 12]. In [11], it is proposed to establish the level of trust of the route using a variety of Internet routers and sender's router with the help of information about the levels of trust of routes that is received as supplementary information from each router on the route through the Internet. Evaluate the explanatory information received via the Internet about the levels of trust to determine the adequacy of the level of trust of a route. When the sender's router receives the route, it determines whether the level of trust is acceptable using the information from the database (DB). If the route is acceptable, it sends a message of confirmation of this route as a code RSVP-TRUST to all nodes of this route. Otherwise, it transmits a message RSVP-TEAR to all nodes of that route, i.e. that the route is not trustworthy. Further, the sender's router continues to choose the route to the recipient router until an acceptable route is found or till when the availability of routes is exhausted. This is done to implement a TR; however, the selection of routes is not great and not very reliable, because, although the number of routes is defined by routing algorithm, yet the route is selected among all the nodes of TN, not excluding untrusted nodes. This reduces the trustworthiness of the route chosen as in it can be node that does not inspire trust. Author in [12] proposed an alternative method of protecting the information without the use of encryption algorithms while the information is being transmitted in distributed networks that are being subjected intentionally to malicious attacks. The author developed client-server application to aid in data transfer over distributed networks, which allows user to transfer data using specific route in order to avoid the traffic flow across zone controlled by an attacker. Of course the method allows to significantly reduce probability of class of active network attacks but within a distributed networks. Also there is a possibility that the attack on trusted servers will be successful. The author, of course, in this case, proposed incorporating multiplexing algorithms, which further ensures the security of information being transmitted. Nonetheless, attacker could still get part of the information being transmitted, if not all.

According to Tiamiyu (2013) as well as Hills (2005), VPN extends private network services for organization(s) across a public or shared network infrastructure such as Internet or service provider backbone network and it has capabilities for tunneling mechanism (encapsulation), based on RFC-1234 [13], support for various protocol and scalability, and may span multiple IP Autonomous Systems (AS) or Service Providers). Also MPLS VPN as a trusted VPN is a high-performance telecommunication network [13, 14] that allows data transfer within a network using labels, which are the information attached to the packet that tells every intermediate router, label-switch router (LSR), to which egress edge label-switched router (E-LSR) it must be forwarded. Also that MPLS is IP-compatible, thus it integrates easily with traditional IP networks. Furthermore that in MPLS, packets are routed along pre-configured label switched paths (LSPs), and packet flows are connection-oriented [13, 15 & 16], and that cryptographic tunneling is not being used in MPLS (like other trusted VPNs) as trusted VPNs rely on the security of a single provider's network for security of the traffic [13].

Considering all mentioned above, VPN method (and by extension, MPLS VPN) is also simple and more effective like the author's proposed method, method of TR but security by encryption and tunneling do not exclude potentially untrusted network or network nodes from participating in the routing of data to ensure a high degree of data security. The privacy afforded by VPNs was only that the communications provider assured the clients that no one else would use the same circuit. This allowed clients to have their own IP addressing and their own security policies. A leased circuit ran through one or more communications switches, any of which could be compromised by someone interested in observing the network traffic.

In VPN, the Internet connectivity and the subsequent tunneling are dependent upon ISP i.e. it is usually impossible for a user to know the paths used by VPNs, or even to validate that a trusted VPN is in place; they must trust their provider completely. And as the data tunnel is rendered across multiple routers, rogue tunnel can be established and an accompanying rogue gateway. Being part of the tunnel, these intermediate rogue routers can examine/copy/modify the data should they be malicious even if the packets are encrypted. In such a case, the sensitive user data may be tunneled to the imposter's gateway. Thus data, which is encrypted between VPN client and the VPN endpoint, can be seen while traversing public networks from sender to receiver [17, 18], and as a result a VPN-based networks implementation can be attacked in many ways. Some of the possible types of VPN attacks are attacks against VPN protocols, cryptanalysis attacks and denial-of -service attacks. And as VPN traffic is often invisible to IDS monitoring when the IDS probe is outside the VPN server, as is often the case, then the IDS cannot see the traffic within the VPN tunnel because it is encrypted. And as such if an intruder gains access to the VPN, the intruder can attack the internal systems without being picked up by the IDS.

Furthermore, in VPN, PPTP and even L2TP are not provided robust security mechanisms against rogue ISPs and tunnel routers. Though the advanced and comprehensive encryption capability provided by IPsec does offer a high degree of protection against these rogue elements yet if node replication which is the process of incrementally copying, or replicating, data that belongs to a client node is in effect, data is replicated and possibly decrypted.

VPN though provides confidentiality such that even if the network traffic is sniffed at the packet level, an intruder would only see encrypted data. But the fact that the encrypted data is available to the intruder is also a big threat as the intruder could one way or the other gets encrypted data decrypted as there exists possibility that encrypted data copied while being transferred over Internet sometimes may eventually be decrypted when the same encryption algorithms then used has become weak encryption algorithm e.g. case study of DES and AES (DES is breakable as the Electronic Frontier Foundation built a DES-cracking machine that can find a DES key in an average of a few days' search [19]. AES is still unbreakable (still unbreakable implies it could be breakable in the future considering the claims by team of researcher as elaborated in [20])). But this is unacceptable for certain confidential information that has to remain confidential now and in the future e.g. certain medical and research information, military information and so on.

According to Bhaji (2008), "a security policy is a set of rules, practices, and procedures dictating how sensitive information is managed, protected, and distributed" [21], thus if security policy by either the sender or the receiver stipulates that the data should not be available to third party or routing devices in certain locations, whether encrypted or not, then data has to be routed securely from sender to receiver in such a way that data is not available to intruder(s), either encrypted or not. And this is of paramount importance as then the intruder has no data to examine/copy/modify, so to say, the intruder has no data to tamper with in any form.

In MPLS either of sender and receiver may not be located outside the MPLS domain because even if the MPLS domain is properly configured, that its internal structure and the core network is not visible to external networks (and thus protected against various attacks, e.g., DoS-attacks), the data are subjected to various attacks outside the MPLS domain i.e. when they cross the border of MPLS-domain. To ensure that data is tranferred reliably and securely through trusted node, MPLS provides ability to control where and how traffic is routed on privately owned network, manage capacity, prioritize different services, and prevent congestion. But this is only possible only within the MPLS-domain.

Thus, deep analysis of VPN (MPLS inclusive) on the subject matter, TR, revealed that basic architecture of *MPLS* network does not provide security services such as encryption, unlike VPN. Therefore, *MPLS* does not protect the *confidentiality* of *data, especially*

*outside its domain [18, 19 & 22]. However,* MPLS or IPsec VPNS, GETVPN (Group Encrypted Transport VPN) and DMVPN (Dynamic Multipoint VPN) use *IPsec,* robust encryption algorithm, to ensure *data confidentiality [22, 23 & 24].* Nonetheless, IPsec relies on the Internet for transport and, thus, third-party can still get copy of encrypted data [25].

### 1.2 Open issue
Though the methods analyzed above are good and acceptable since, to some extent, they provide for TR. However, the open issue is still how to prevent third party from having copy of the data being transmitted, encrypted or not. Since having a copy of the data in encrypted form does not guarantee that the third party would not be able to decrypt the data. It could be a matter of time. Also the third party may play other known and unknown pranks on the data being transferred. This issue is being addressed in the implementation of TraConDa that is based on trusted routing mechanism by preventing third party from having copy of the data being transferred. Since no copy implies nothing to be decrypted, and no access -nothing to play pranks on.

Moreover, if the data is properly encrypted, how can one be sure that solution has not been gotten to that encryption method/mechanism already though it is believed to be most reliable security method/mechanism? Maybe it is just yet to be made a public knowledge (because of personal interest of developer to spy on others or national policy/interest). As explained earlier on, present strong encryption can be a weak encryption later as a result of technological advancement and eventually encrypted data could be decrypted and made a public knowledge that could result into many things among which is war if it is military secret, bridge of contract (if it is medical record not meant for public knowledge).

So, what is the essence of encryption if the encrypted data could be decrypted later, because of technological advancement, which is very possible in these modern days? And as one can only decrypt, if possible, only what is accessed or accessible; no data or no access to data implies nothing to decrypt either now or later.

As a result, architecture and analysis of requirements for TraConDa is paramount. Thus, TraConDa implementation stages are subsequently described.

### 1.3 Implementation stages of TraConDa

Having analyzed various security mechanisms and some of other works done by researching community worldwide, especially MPLS VPN, the conclusion is that TR mechanism procedures to guarantee confidentiality of data transfer over GDTN like Internet should comprise the following stages:

- **defining topology of network that is participating in information transfer**

Topology is defined at initial stage and it entails gathering route information (for example, with the help of traceroute using IP-option and utility like icmp) and routing table information (by accessing routing table via SNMP or BGP protocol), prepare topology map based on the information gathered, resolve problems/conflicts associated with IP-aliases (many works are done on mapping topology of networks, e.g. Internet, among which are works done by authors in [26, 27 & 28];

- **extended identification of RD**

Extended node identification process entails gathering information and identifying possibility of remotely getting access and controlling any of devices within the map created, and subsequently converting them to trusted routing device (RDt) when necessary and where possible (conversion to RDt could as well be achieved by introducing program module into OS of RD following the work done by authors in [29, 30 & 31]). In the interest of extended node identification could be used WHOIS server (for georeferencing), traceroute utility, Nmap port scanning as well as SNMP traffic interception. For remote access to TN nodes can also use SNMP brute force attack on password, HTTP-vulnerabilities exploitation, brute forcing a Telnet password, DHCP-service exploitation, SNMP-vulnerability exploitation and so on.

- **defining trusted route**

This is done by finding some optimal and shortest active routes from sender to recipient within the map using algorithms by authors in [321] and save the active routes found into DB. Thereafter, the RD within those active routes are converted to RDt, if possible, after gaining remote access to them. Then trusted route is either calculated using modified Dijkstra's algorithm [33], the algorithm that finds trusted shortest path (the path that comprises of only RDt as described by Tiamiyu (2014). Alternatively trusted route could be selected analytically by finding at least a route that passes only through a set of RDt from a packet sending point (initial IP-address) to a packet receiving point (destination IP-address);

- **forced routing (traffic control) and monitoring of regenerated network**

This stage is about redirecting traffic along trusted route, i.e. path comprising only RDt (by source routing, for instance), and monitoring status of RDt in the interest of TR mechanism.

All these stages of TR mechanism are further described in details in the subsequent chapters.

## 2. Trusted routing mechanism architectural requirements and implementation algorithms

On architecture of execution, the most efficient is a two-level scheme for the construction of TraConDa, which includes TR manager (TRM) and TR agent (TRA) for traffic control (TC) and BCCH (backup control channel) as their interacting interface. TRM is installed to automated workstation (AW) where data traffic control takes place in the interest of TR. TRA are implanted to some RD participating in data transfer within the network to which AW is connected. The modified OS image i.e. OS image with TRA embedded as program module (PM) can be copied to RD using tftp protocol [34].

Also there is possibility of recording any code and switching OS to execute it while RD is still operational, which provides the possibility of realizing extra control mechanisms of RD. Such a possibility of changing the OS code during operation allows carrying out automated embedding by connecting PM straightaway (without modification and packaging of OS image file, load it via tftp and reboot RD). More details on the embedding TRA and other components within structure of TraConDa is subsequently discussed in details in Section 2.1. Modular structure of TraConDa is shown in Fig. 2.

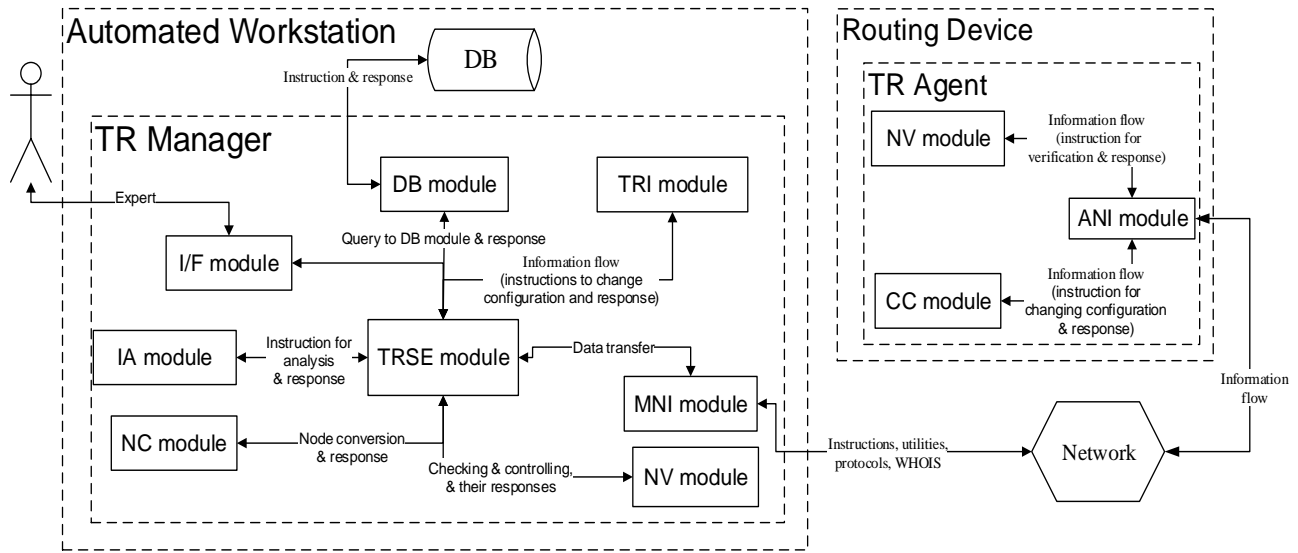

**Figure 2: Modular structure of TraConDa**

**2.1. Trusted routing agent embedding as a functional requirement in TraConDa**

Embedding TRA as a functional requirement in TraConDa entails getting remote access and control to nodes in TN. And subsequently embed PM in the form of TRA to OS of the nodes. When embedding PM, according to [31], compressed OS image is the first to be disassembled and then analyze its content to find out how and where to insert the PM and thereafter embed codes, repackage it so that it can be installed back to the RD. And repackaging includes corrections to checksum of the modified OS image (if necessary) and adjusting header (for ELF - header and sections, and for MZIP - header and checksum recounting) so that it can pass through the startup tests, which may prohibit the modified OS image file from running on the RD when the checksum is not correct.

Considering the works of the authors in [29, 30 & 31], developing and embedding PM should be carried out as follows:

**1. Identify and detect OS module version of the RD;**
OS image file can be downloaded from RD using FTP or TFTP server.

**2. Disassemble and analyze the OS image in the IDA Pro Disassembler;**
When carrying out the analysis, first thing to do is to detect and identify functions in disassembled OS image. In the work by the author of [31], IDA pro was used.

**3. Analyze disassembled OS image further using IDApython and correct errors allowed by IDA pro Disassembler in the functions allocation;**
The author of [31] used IDApython to further analyze disassembled OS image to get critical information for the survival of PM to be embedded. Since several functions and links would not be found using only IDA Pro, the author used IDAPython in the recognition process of functions and strings as well.

**4. Develop PM to be embedded;**
OS is written in C (with low-level assembler insertion in the kernel), linked with the help of compiler *gcc* (with certain modifications, for example, scripts for assembling) and has a modular design [30]. Thus, the idea follows that PM can, likewise, be developed for embedding also in C language to obtain machine codes that should be linked to the original OS image, i.e. copy to the selected inactive memory area.

**5. Select area in the disassembled OS image where PM will be embedded;**
There are three areas suitable for embedding PM, which are free levelling area (area between the data and code segments), area in a code segment and area in a data segment [30]. The author in [31] choose data segment to place the code, sacrificing debug line, which is almost, probably, will never be used as there are a lot of such strings in OS of RD, for example, in all versions of Cisco IOS. The author further decided to write a NULL (zero) character to the first symbol to prevent the string from causing problem, as well as to avoid suspicion from the user, if the system administrator decides to use some functions of OS, which require that line.

### 6. Prepare the selected area for embedding process;
As necessary, make changes that are needed for preparation («deactivation») of selected memory area, for example, in the case of using strings, its truncation [30]).

### 7. Embed PM into the selected and already prepared area of the disassembled OS image file;
Identifying offset of the selected area in the disassembled OS image file and placing segment(s) of PM into the field.

### 8. Connect the embedded PM;

This is done by replacing instructions (function calls) in the modified OS image file, with instructions (function calls) of the PM. This requires the replacement of transition address (or offset) in the function call instructions. Setting up the right calls inside PM, as well as OS function calls from the PM. Connecting modules by calls transfer method requires changing the call instruction and depends on the architecture of processor. For instance, for MIPS processors, value of the call instruction is low 28 bits of the address of the called function, shifted by two bits to the right while for PowerPC processors, value of call instruction is 26-bit difference (with sign) between the address of the called function and the call instruction.

### 9. Repackage the modified OS image.
This should include corrections to header of packaged OS image. So also correction to checksum of the modified OS image (if necessary). After all these, the modified OS image can be installed to RD and tested. Flow-chart of processes involved in embedding TRA into OS of RD (RD to RDt Conversion) is shown in Fig. 3.

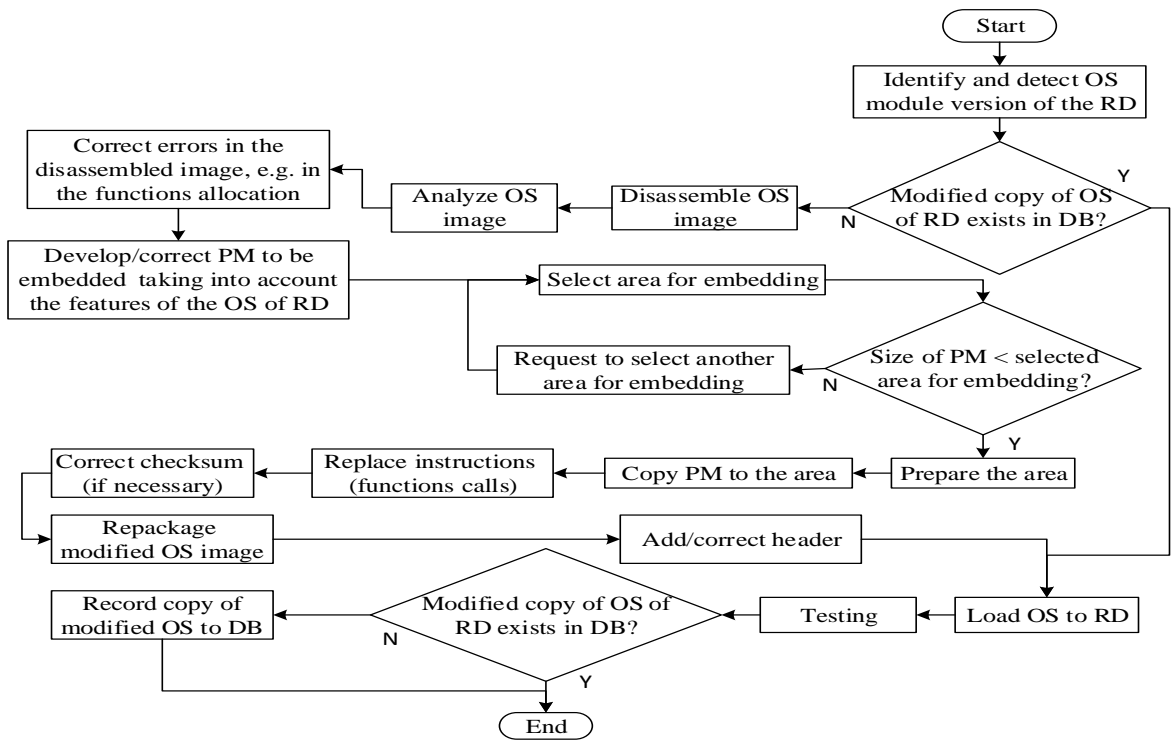

**Figure 3: TRA into OS of RD embedding process (RD to RDt conversion process)**

## 2.2. TraConDa operation algorithms

Getting TraConDa for traffic control (TC) in TN necessitates development of algorithms for operations of TRA and TRM.

## 2.3. Trusted routing manager operational designation and method

TRM for TC is designed to create secured communication channel for TRA remote control; exercises control over functioning of the TRA (i.e. after embedding, PM allows to have full control over RD as then the control is transferred from OS of RD to the PM. Thus the RD is remotely control by TRM, and as TRA carries out instructions contained in special packets via BCCH from TRM, TRM controls functioning of TRA); generates and transmits control commands to TRA. Also it is designed to gather information about routes and RT of routers as well as identifying and obtain specific (active) route of traffic. Furthermore, it is designed to implement georeferencing, TC and displaying or printing of results (Fig. 2).

TRM accomplishes some of these tasks by adding or changing information about the composition and characteristics of the transit routers and the functioning of TRA whenever necessary and getting responses from TRA in form of messages containing information about results of executing the command. For program realization of the required tasks of TRM, the following modules are being utilized. Trusted routing stages execution (TRSE) module. It is main module in TRM and it provides for interaction of other modules by supporting among others initialization, synchronization, execution and monitoring of their algorithms. TRSE is managed via interaction interface (I/F) module through which expert launches TraConDa and performs basic control (for example, login, shutdown etc.). Manager's network interface (MNI) is being used to provide interface with network as well for data encryption (decryption) and launching of network utilities as well as sending instructions to TRA. Database (DB) module renders high-level data access interface to other modules (for example, getting IP-address of trusted route among others). Maintain channel in trusted state by verifying and monitoring nodes is the purpose of node verification (NV) module. Node controlling (NC) module is for conversion of node to trusted entity. And for process of routing of data along specific route is used trusted routing implementation (TRI) module. Information analysis (IA) module is for analytical processing of information.

DB is local to AW, and it stores information that is used in TraConDa. Modular structure of TRM as part of TraConDa is shown in Fig. 2.

TRSE module essentially manages the DB module, MNI module, IA module which can be regarded as auxiliary within the structure of TraConDa. DB module interacts with the DB by means of instructions. It provides data only on request.

NV module realizes monitoring of system status by carrying out periodically extended node identification along trusted channel using MNI module. Whenever untrusted node is found along route of data transfer, NV module signals to TRSE module about this, which brings about an instantaneous usage cessation of the trusted path and possibility of search for another. Thus, NV module works in parallel with TRSE module in the background and it is considered a «TRM demon».

By request from TRSE module, NC module initiates process of converting node to trusted entity. For this, it uses a formalized information about node and generates a request for its modified OS from DB, if available. In response, the module receives the required OS and loads it to the RD (hardware node) by establishing automated remote access to the node, for example, using FTP (TFTP) (whenever this operation is successful, RD becomes RDt). Also, by the request from TRSE module, TRI module initiates the process of routing confidential information along given trusted route. TRI module, for this, receives information about trusted route from DB and redirects information along the route, (the process of forced routing).

In normal mode, TRM expects command from the administrator of AW and listens to BCCH while awaiting messages from TRA. When commands are received, TRM enters regime of command execution. Information about route or RT received from TRA are stored in DB after (if necessary) preprocessing like formalizing RT view. The followings are some of commands TRM executes:

## Command № 1 – Get RT using SNMP

By command 1, RT of a specific router is received from specified TRA via SNMP routing protocol. The processes involved are select the desired router from list of available routers; generates necessary command to get RT via SNMP protocol (Fig. 4); and verify availability of the received RT in the DB. Access to RT of the specified router via SNMP control protocol is implemented within command execution block (Fig. 2) within TRSE module by activating block that implements access method via SNMP. Routes are MIB objects. According to the MIB requirements, each route in the DB of router corresponds to a record [35]. TRM generates to MIB standard SNMP-request according to the standard procedure of SNMP protocol, (1.3.6.1.2.1.4.21–ipRouteTable and 1.3.6.1.2.1.4.21.1–ipRouteEntry) where iso(1), identified-organization(3), dod(6), internet(1), mgmt(2), mib-2(1), ip(4), ipRouteTable(21) and ipRouteEntry(1) [36, 37]. On getting a response, TRM then connects with the specified router via SNMP and starts recording information about RT from the router. On starting, TRM receives response from the specified router with information about the destination, ‹ipRouteDest›, and then sends ‹get-next› request to the router to get the next parameter: interface index ‹ipRouteIfIndex›. By sequentially sending request ‹get-next› and receiving response from the router, TRM receives detailed information of all the records (for example, IP-address destination, destination interface, next-hop, protocol code, route mask etc.) that are located in the RT of the specified router. Detailed information about RT of the router is recorded into DB for further analysis and for preparation of active route(s). This cycle is repeated for other routers [37, 38]. Flow-chart of processes involved in this method, as a result of executing command № 1 (Get RT using SNMP), is shown in Fig. 4.

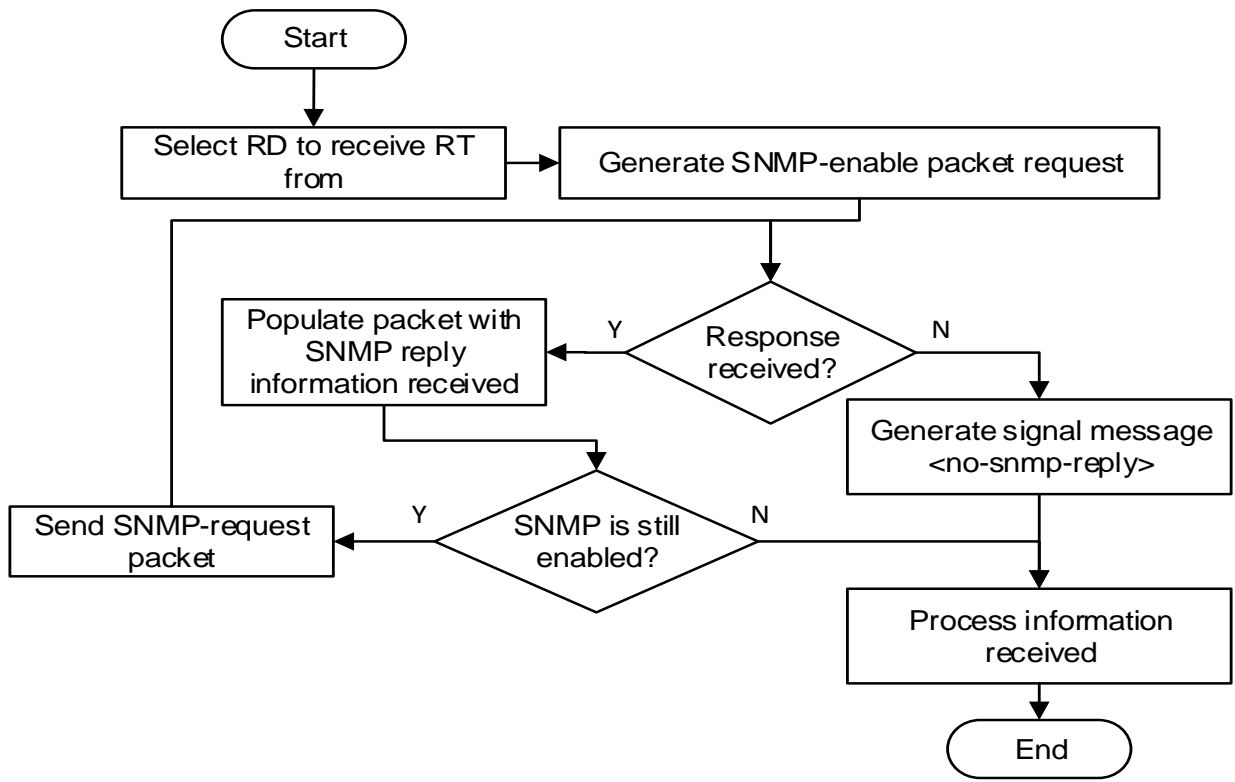

**Figure 4: Getting RT in TraConDa using SNMP**

**Command № 2 – Get RT using BGP**

By command 2, RT of specified router is received via BGP routing protocol (Fig. 5). The processes involved are similar to that of getting RT using SNMP protocol but here, BGP-session is opened with the selected router instead of enabling SNMP. And the BGP-session is broken with the selected router after receiving the RT or when keep-alive signal stops. Whenever BGP-session is established with a specified router successfully, the router begins to share its RT with TRM. RT information exchange with TRM begins with the help of packet «update». On the TRM, the following control fields are filled: IP - address of the router; Version – BGP version; AS - autonomous system; hold time - 180. Then TRM effects primary assessment and analysis of RT information contained in the packets «update» and generates information obtained (information like Host BGP IP, Unfeasible Routes, Length NextHop, Next Hop, Length ASPPath, ASPPath, Networks etc.) [39], and sends it to DB. Flow-chart of processes involved in this method, as a result of executing command № 2 (Get RT using BGP), is shown in Fig. 5.

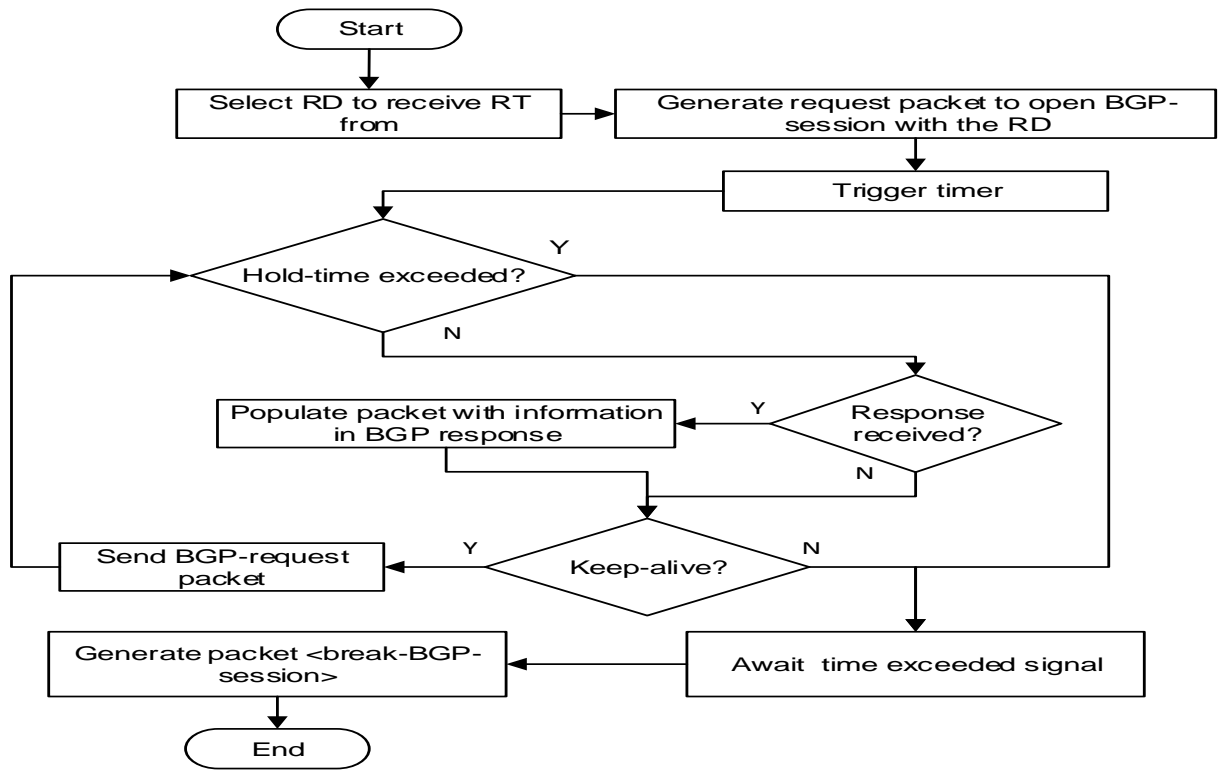

**Figure 5: Getting RT in TraConDa using BGP**

TRA can simultaneously support BGP-session with multiple routers and this allows receiving in real-time RT of different routers or changes occurring within them [39].

**Command № 3 – Get route using IP-option**
By command 3 TRM implements tracing of data flow between two specified RD using parameter IP option (Fig. 6). This process involves selecting a desired or specific RD as destination using DB where list of active routes among others are stored, and then generates necessary command to trace packet route (Fig. 4). After receiving the results of the trace, analyze the received route of information flow and verify availability of the received route in the DB. This signal from TRM to get route down to a specified IP-address of a specific RD using IP-option is based on the parameter «record route» option field of IP-packet and it is designed to obtain information about the active route of the information flow (traffic), i.e. list of IP-addresses of all transit routers down to the specified router, in between the sender ( and recipient, another remote but specified RD (remote RD_x, where x is 1 to $n^{th}$ number of RD. e.g. RD_1, RD_2…RD_n). TRM as sender includes in the packet parameter «record route», thus, IP-address of every transit router participating in the delivery of the packet is stored in the packet.

Therefore, along with the packet, RD_x as recipient receives a list of all transit routers «visited» by the packet and then generates and sends response packet containing list of IP-addresses of all these transit routers and that of RD_x. Having received the response packet from RD_x, TRM effects preprocessing of the data about traffic flow and sends received information to DB. Flow-chart of processes involved in this method, as a result of executing command № 3 (Get route using IP-option), is shown in Fig. 6.

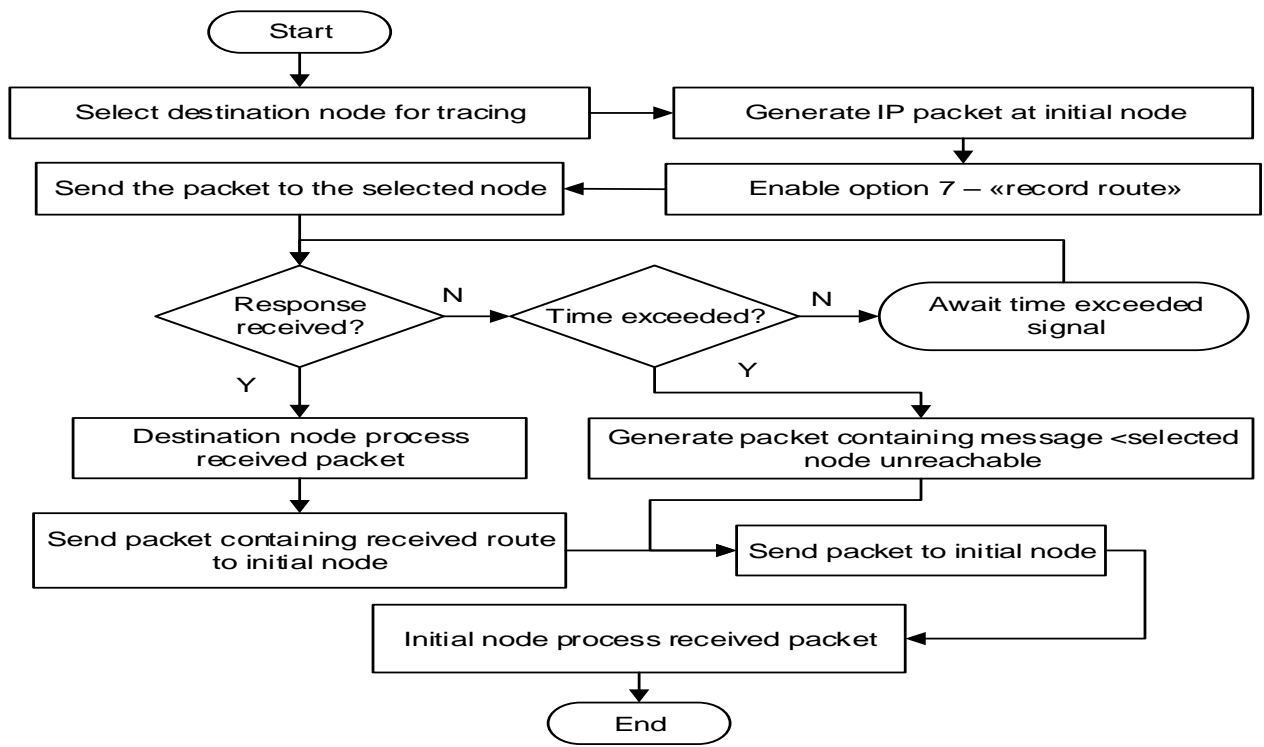

**Figure 6: Packet Tracing algorithm using IP-option parameter**

**Command 4 – Get route using ICMP-request**

By command 4 is implemented tracing of data flow using ICMP-request to a specific RD with a given IP-address (Fig. 7) and this necessitates, in sequence, selecting a desired RD, and then generates necessary command to trace packet route. Thereafter analyze the received route of information flow after receiving the results of the trace, and finally, verify availability of the received route in the DB. Whenever signal comes from TRM requesting route to a specific IP address using ICMP-request, within command execution block in TRSE module (Figure 2), as usual, is activated corresponding block that implements the method of tracing data flow using ICMP-request.

Basis of tracing data flow using ICMP-request is the use of field value TTL (Time to Live - the lifetime of the transmission of packet) located in the IP-header and ICMP-message ‹Time Exceeded›. Two techniques being used to trace packet route to the destination are small values of TTL and invalid port number (for example, 33434) [40].

For the detection of intermediate routers between TRM, for instance, and the specified router, ICMP-request packet with a small lifespan is sent. The counter value starts at 1 and increases by 1 for each group of three ICMP packets. Having received a packet, router decreases TTL value by 1. Whenever TTL=0, packet is not sent further and TRM as the sender receives ICMP message ‹Time Exceeded›. TRM remembers one line of output data for each router from which the message ‹Time Exceeded› was received. When the receiving host i.e. the specified router receives ICMP-request packet, it returns an ICMP-message ‹Unreachable Port›, indicating that the packet reached the destination, thus, end of the trace. This happens because TRM intentionally uses an invalid port number to force an error. TRM builds a list of intermediate routers, starting from a distance of one transition and ending with the destination host and sends it to DB. Flow-chart of processes involved in this method, as a result of executing command № 4 (Get route using ICMP-request), is shown in Fig. 7.

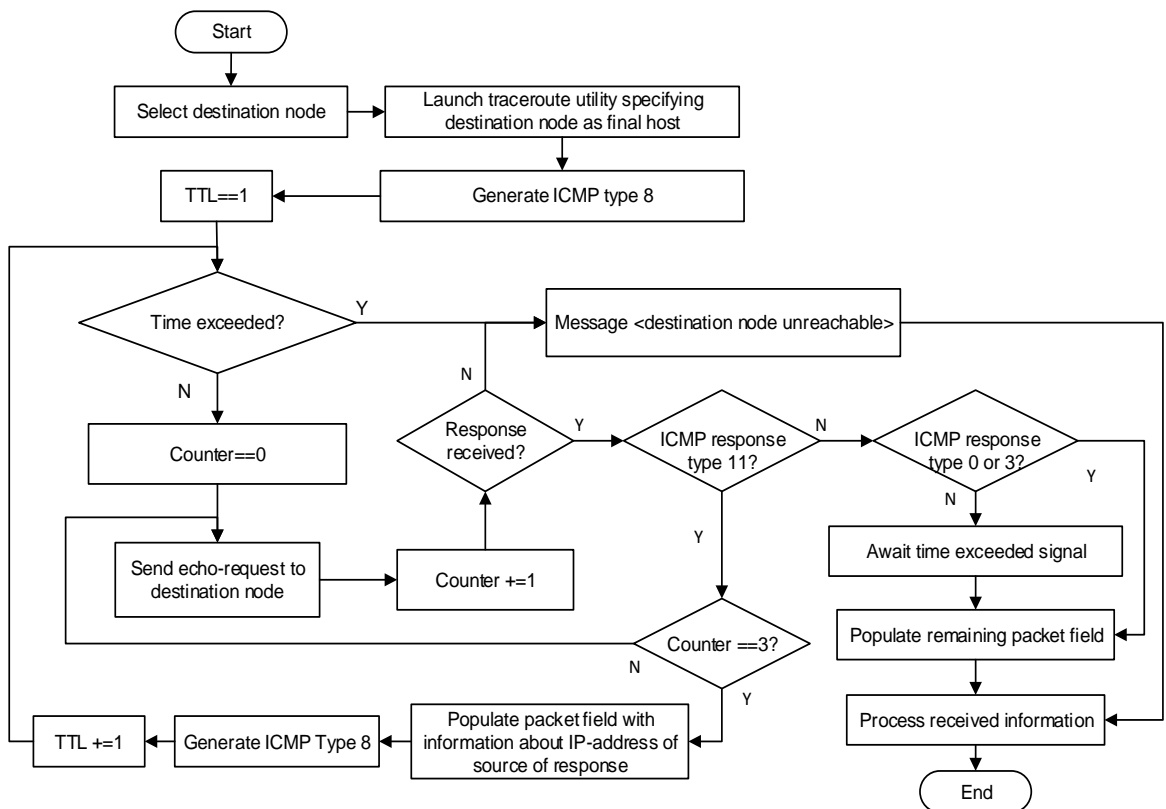

**Figure 7: Packet Tracing algorithm using ICMP-request**

Though the TTL field of IPv4 has been renamed to Hop Limit in IPv6, the message format, basically, is the same for ICMPv4 and ICMPv6. Regardless of the name, the field still has the same basic purpose of keeping a datagram from wandering the Internet forever [41].

**Command 5 – Get trusted route**
By command 5 is implemented process of getting trusted route by tracing of data flow between two points on a topology map created during topology definition. The processes involved are analysis of formalized RT in DB in order to obtain route of a specific packet by selecting starting point IP-address (fixing starting for the route) and destination IP-address, and then identify a row in the formalized RT using the selected starting point IP-address; then analyzing RT, precisely by the starting point of the route, IP-address (Next hop) of other transit routers to the destination IP-address is determined, thus search for active IP-addresses of intermediate routers is performed iteratively, which gives route of a particular packet, which is then recorded into the DB as active route. After active routes are defined, conversion of RD to RDt (Fig. 3) (as many as possible, if not all) within the active routes are done. Then, trusted route is gotten by iteratively finding a route that passes only through set of RDt from a given starting point to a destination point. This information about trusted route(s) is stored in DB for traffic redirection in the interest of TR from that given point to the destination point.

After conversion of RD in active routes to RDt, for accuracy and time-efficiency, trusted route is calculated using modified Dijkstra's algorithm [33]. Practically, it corresponds to the basic Dijkstra's algorithms with the exception that it checks trustworthiness of every next node (possible next-hop node) while finding shortest path before including such node in the process of finding shortest path. Any node that is not trusted is avoided and excluded in the process. Flow-chart of complex processes of getting trusted route using modified Dijkstra's algorithm is shown in Fig. 8.

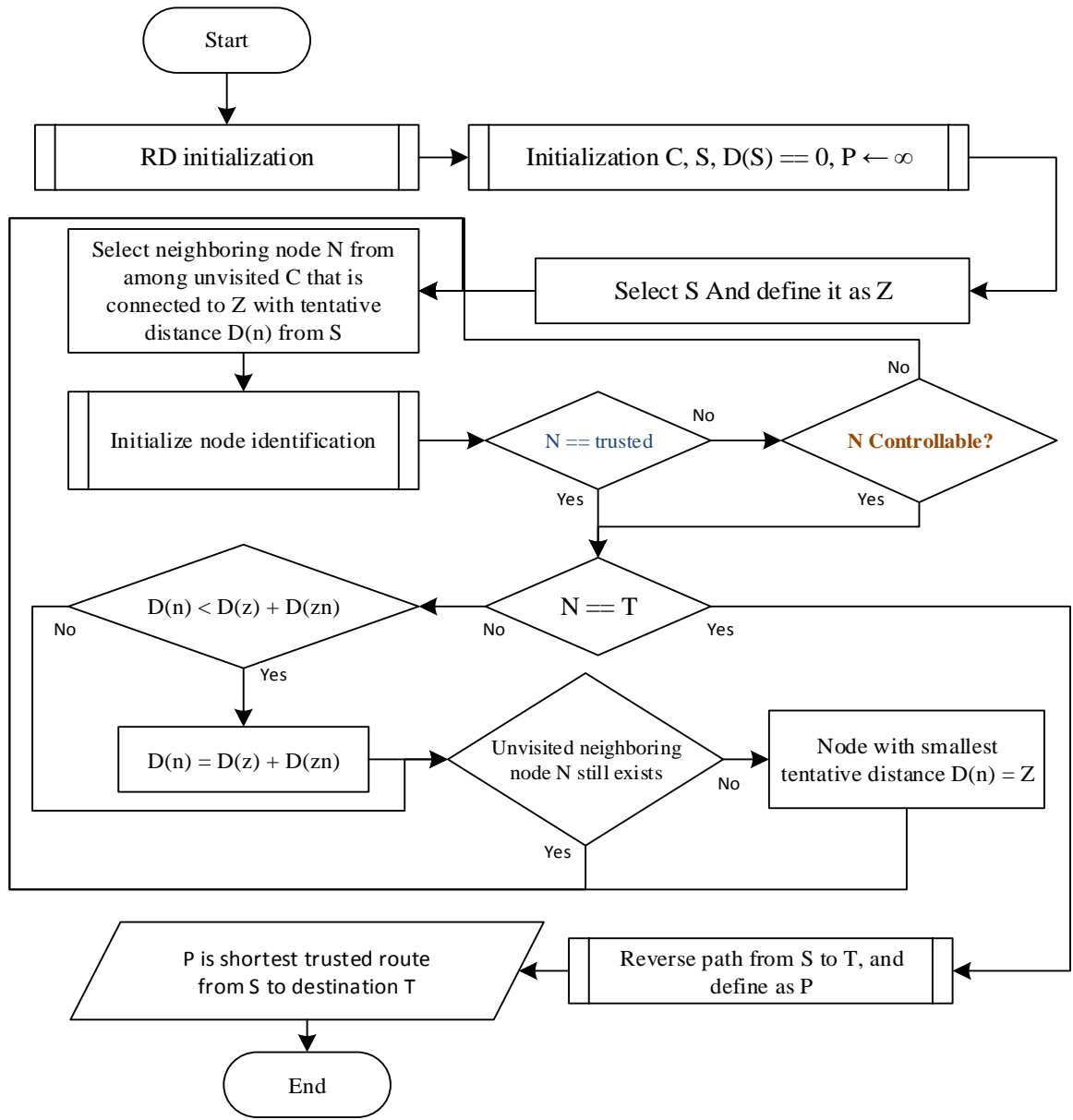

**Figure 8: Modified Dijkstra's Algorithm**

As noted in Fig. 8, RD are set of nodes (vertices) in a network, and they are marked as unvisited set C. D is tentative distance value, S is initial node and D(S) == 0 (tentative distance value for initial node S is set it to zero). P is shortest path, initially P←∞ as path is not yet found or determined. D(n) is tentative distance between current (neighboring) node under consideration to the initial node S; D(z) − smallest known tentative distance from Z to the node S, and D(zn) − smallest known tentative distance from the neighboring node N to Z.

**Algorithm works as follow:**
1. Assign to every node a tentative distance value: set it to zero for initial node (i.e. D(S)==0 and to infinity for all other nodes (P←∞);
2. Set the initial node, S, as current Z. Mark all other nodes unvisited. Create a set of all the unvisited nodes called unvisited set C;
3. Select neighboring node N from among unvisited C that is connected to initial node S;
4. For each neighboring node N with tentative distance D(n), cross-check trustworthiness of the node. Every node N that is not trusted is excluded and marked as visited;
5. For each neighboring but yet unvisited nodes N that is trusted (i.e. N==Trusted) and that is not the destination node (N!=T), calculate its tentative distance (D(n)=D(z)+D(zn)). Compare the newly calculated tentative distance to the current assigned value and assign the smaller one (i.e. compare if D(n) < D(z)+D(zn));
6. When all of the neighbors of the current node are fully considered, the current node Z is marked as visited, and it is removed from the unvisited set C;
7. The visited node with smallest tentative distances D(n) is afterward considered as new current node Z;
8. Define path from the new current node Z to S as shortest trusted route P from Z to S;
9. Repeat steps 3 to 8 until the target node T (destination node) is reached, i.e until N==T;
10. Reverse path P found (when N==T). Then P is shortest trusted route from initial destination S to the target destination T.

It is worth noted that according Tiamiyu (2014), using modified Dijkstra's algorithm, trustworthiness of every RD in a given network is determined in accordance with the formula described below:

$$N_i \in Q, \qquad\qquad 1$$
и

$$Q = \{a, b, c, d\} \qquad\qquad 2$$

where $N_i$ – network node being cross-checked;
Q – set of parameters for which the node is checked;
a, b, c, d – parameters which define the node trustworthiness,
and

a − determines authentication that node is able to perform in the process of verification of its physical location, as the node could be trusted due to its particular geographical location;

b – identifies a node known to sender router as untrusted due to country or corporation in which the node is located;

c − determines the average of the measurements of the physical location of the node obtained by "pinging" the network. The physical location of the node can be determined by placing it in a certain secured place, for example, in a military base or government building;

d − characterize the information found in the node that is being stored there for the purpose of identification, for example, a software module in the form of agent.

In TraConDA, for time-efficiency also, active route is found using other applicable algorithm, precisely X-BDV algorithm described by authors in [32]. A recent study (Abraham et al, 2013) showed that X-BDV algorithm that applies a bidirectional Dijkstra's algorithm (BD) for target and exploration queries, and X-CHV that runs on top of contraction hierarchies does look for routes that are substantially different from the shortest path and are locally optimal. The work actually was based on finding multiple routes between two vertices on a directed graph with nonnegative, integral weights on edges. And its advantage over other algorithm being used to find alternate routes e.g. choice routing algorithm [42] is that it is a faster algorithm for continental-sized networks. Unlike choice routing algorithm, X-BDV is successful more often (compared to other algorithms of its kind). Experiments carried out in [32] showed that the algorithm, X-BDV (simplified version of BD) is practical for real, continental-sized networks, and it can be used to find multiple alternative paths.

Thus, in TraConDa, X-BDV algorithm is used to find certain number of alternative shortest routes from sender of confidential data to the recipient of the same instead of finding all at once. This could reduce time to find trusted route considerably as there is possibility of getting trusted route from among first few active routes found (thus reducing the number of RD to be converted to RDt). The number of alternative shortest routes can be increased accordingly, especially when necessary i.e. when trusted route is yet to be found from the sender to the recipient.

Flow-chart of complex processes of getting trusted route using X-BDV and modified Dijksra's algorithms is shown in Fig. 9.

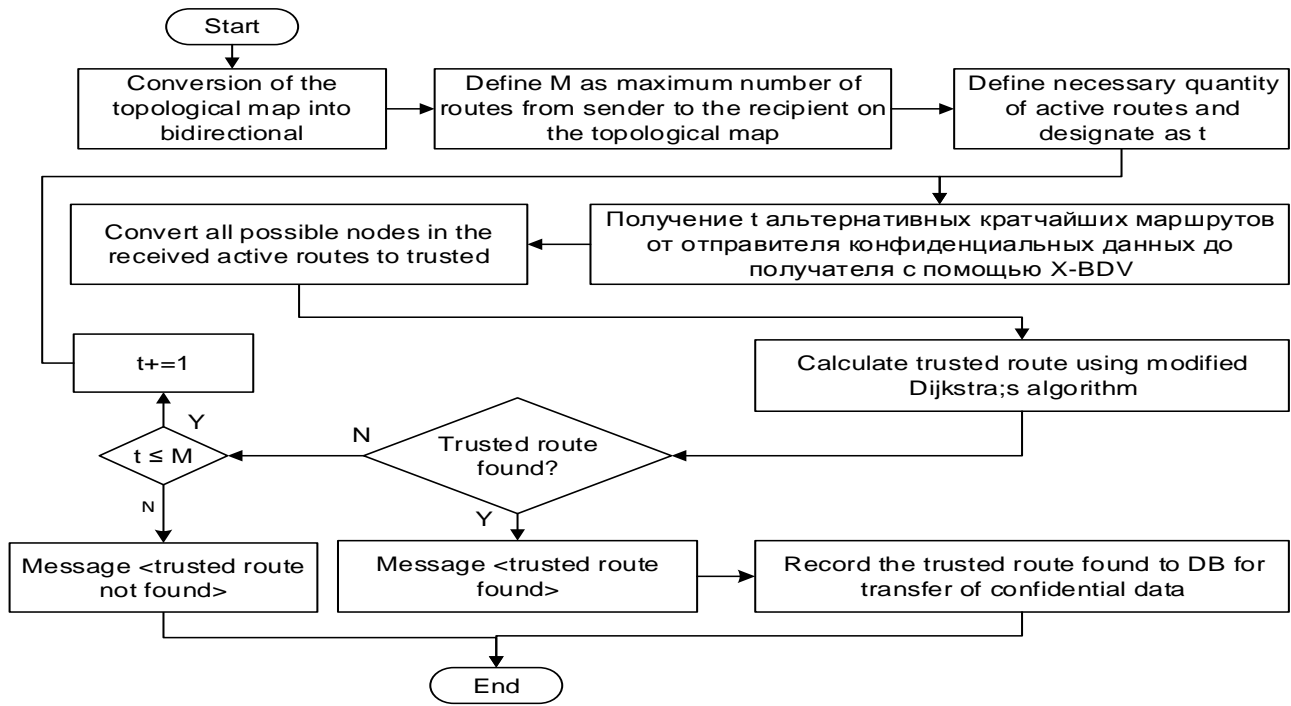

**Figure 9: Getting trusted routes using X-BDV and Modified Dijkstra's Algorithms**

Processing of the following messages from TRA by TRM are always allowed even in standby mode:

i. Acknowledgement from TRA about receipt of command from TRM. Upon receiving such messages, TRM, by code of the message, determines the semantic meaning of the message and generates a text string that characterizes the incoming message.

ii. Informational messages from TRA about routes of transmission of messages, determined using ICMP-request or IP-option parameter «record route». Upon receipt of such messages, TRM, by code and the notational content of the message, determines the beginning and the end of the route, the list of transit routers included in the route.

iii. Informational messages from TRA about router's RT, received using BGP or SNMP protocol. Upon receipt of such messages, TRM, by code and the notational content of the message, determines router from which the RT is received, and puts the received RT into the DB.

In the mode of operations with the DB, it is required of TRM to:

i. Add new TRA into the DB, modify the parameters of selected TRA, remove selected TRA from the DB, add additional DB entry point for the use of SNMP or BGP when accessing the required router for the selected TRA;

ii. Add new router to the DB, change parameter of selected router, remove selected router from the DB.

### 2.4. Trusted routing agent's operational designation and method

TRA is designed to work in public networks like Internet, and in it is realized function of receiving control signals from TRM and decrypting them. Whenever TRA starts operation, control is passed to the block that implements the operation required as depicted in the command. Results of the operation are transmitted to the input of the block for preprocessing and subsequently to the block for sending of results and acknowledgment when it is appropriate.

TRA receives control signals from TRM via BCCH (Fig. 2) and executes instructions contained therein. In addition, ANI module, a component in TRA, helps in traffic redirection by intercepting incoming packets to RD, check their compliance with the requirements and, if necessary, their forwarding. The preprocessed results and information received are sent to TRM via BCCH.

For program realization of the required tasks of TRA, the following modules are being utilized (Fig. 2). A basic module for verification of RDt (NV module) for self-test of agent, verification of OS stability and correctness of node's data. Also as in the TRM, a network module (ANI module), which, in addition to decrypt network packets, can process and single out special instruction, determine information within the instruction and generate response to the instruction. CC module is being used whenever there is necessity to change configuration of node (for example, adding or deleting static route; enabling or disabling certain services, source routing for example; load or reload OS (modified or factory version). Scheme of modules of TRA, including all internal and external interaction, is shown in Fig. 2.

In normal mode, TRA listens to BCCH, awaiting commands from TRM. On receiving signal, ANI module decrypts and analyzes the signal. In the case of nonconformity, signal of non-compliance is sent to TRM, otherwise the command is executed. Normally ANI, based on analysis of network packets, does single out instructions intended for it, decrypts it (if necessary) and sends them to one of the modules:

− NV module, if requested for node verification;

− CC module, if changes to node configuration is required;

thus the followings are types of command TRA executes:

i. Command № 1 – Get node status
General information about status of each node is collected periodically by TRM using probing algorithm for extended identification and status monitoring for changes that are not initiated by TRM. Details about probing algorithm as a solution to traffic flow control challenges is given in section 4.

ii. Command № 2 – Modify node configuration

By this command, TRM sends instruction to effect specific changes to configuration of a specified RDt, For example, information about static routes has to be removed each time transmission of confidential information is complete so as to bring back the network to initial stage in order to maintain secrecy of operation of TraConDa.

Having received such instruction and analyzed it in ANI module, TRA passes control to CC module, which will make the necessary changes to the RDt. TRA confirms successful execution of instruction by composing and sending applicable response via ANI module. The response is transferred via BCCH to TRSE module in TRM where it is analyzed and then entered into DB.

## 3. Trusted routing mechanism requirements and challenges

### 3.1. Structural and implementation challenges

TraConDa is composed of modules such as TRM and TRA which in turn are comprised of components like MNI and TRI modules. TRM resides in management AW. Thus AW manages RD via BCCH that connects TRM and ANI module. This control channel provides two-way communications. Harmonizing MNI and ANI modules requires interaction interface, i.e. strict communication protocol. And this communication should be carried out by packets that are sent over data networks. Developing a protocol requires defining packet format and message types. And since the packets would pass through different network devices, their format must be correct and conform to generally accepted standards. In other words, these packets must be correct IP-packets (otherwise, they would not just get to RD). Therefore, first of all, it is necessary to select the type of IP-packets to be used in BCCH, i.e. define higher-level protocol that would be used. Basically, RD can forward any valid IP-packet regardless of its type, and the data contained therein. However, in practice, access lists might be used, limiting the passage of certain packets. In addition, packages of rare, obsolete or undefined types can bring suspicion. Other types of packets (e.g., TCP) may be tested for the presence of link or correctness of the header of higher level (for example, TCP could be filtered by port numbers) that requires the correctness of this header in the packet. Thus, packet type must be common, simple as possible, and not arousing suspicion.

Also it is very necessary that traffic be monitored to ensure that data are being transferred accordingly, and in case of unwanted changes or actions of any kind, necessary measures are in place to maintain data integrity and confidentiality. TraConDa is to achieve these goals among others. Thus for the purpose of achieving these, TraConDa has modular structure that comprises TRA and TRM which have BCCH as interacting interface. More details about these modules of TraConDa and other components within them in relation to challenges of implementing TraConDa are described further.

### 3.2. Trusted routing agent's requirements and challenges

TRA allows for a proper management of RD from within. Since RD is controlled by the OS operating system, in which traffic control is implemented. For implementation of traffic redirection in the interest of TR, it is a necessity to implant PM, TRA, in OS so that the integrity and confidentiality of the required traffic redirection is enforced/provided/monitored by this embedded PM. Process of modifying OS of RD to implant TRA requires a substantial timeframe (see Section 2.1 for details of this process), thus, it is carried out in advance, and successfully modified and tested OS is stored in DB which is then available during TraConDa operation. More about TRA in section 2.4.

TRA must aid traffic redirection and be able to read and write to memory of RDt. Thus, CC module, a component in TRA, allows to read and write required data from OS memory, to make changes to the working of RD (though this is possible only if the RD allows reading and writing to its memory (see section 2.1 for details on how to make RD a trusted entity i.e.

have full control (administrative access) after gaining remote access to it). In addition, CC helps in traffic redirection by adding/modifying static route of RDt based on instruction from TRM. Singling out instruction from incoming packets to RD is a necessity in TraConDa, and this is why ANI intercepts incoming packets, check their compliance with the requirements and, if necessary, their forwarding to applicable module.

### 3.3. BCCH operational challenges and instruction/protocol format

Standard control channels (via the console, remote telnet or configuration file) have several disadvantages as they may not be accessible when there is a change in the authentication system (e.g., changing of password or access level); also fact about their control can be detected easily; furthermore, they are limited in functionality since not intended for control of undeclared OS modules. All these lead to the need for BCCH that directly provides for remote control of the embedded TRA modules. As interaction interface requires strict communication protocol, most preferred protocol for BCCH is the use of ICMP protocol. This is because among widely used three types of IP-packets that are TCP, UDP and ICMP, the first two are more complex and are used to transmit data of higher levels, thus, are more likely to be filtered or checked. In contrast, the ICMP protocol is used for controlling and determination of faults [43], so not usually seen. Its flow is usually not restricted (except for closed network, where all external packets could be filtered). Moreover, ICMP packets also helps identify the system as a router when router responds to ICMP requests. ICMP message types are "echo requests" and "echo replies"; a mode which is very like the way a TRA module, ANI, operates when responding to requests from another module, MNI for instance, via BCCH. ICMP echo request and echo response are used in common network testing program, *ping,* to determine packet transmission from one node to another. And data requests, usually, do not arouse suspicion, especially those directed towards RD. In essence, basis of interaction interface via BCCH using ICMP protocol is "adequate". And for messages from TRM to TRCM would be used echo requests, and echo reply for responses.

This is achievable as, according to standard, ICMP-packets consist of a header and data [44]; and RD, when receiving an echo request, changes incoming and outgoing addresses in the request, and then sends back echo reply [45], i.e. not drawn to the data field. Thus, organizing BCCH transmitted data can be resolved in the data field of ICMP-packet without any restrictions (except IP-packet size). In case of sending such a packet to RD that contains no embedded modules i.e. where TRA is not resident, this packet will be processed just like an echo request and to it is returned echo reply. However, whenever response to this special packet, ICMP-request with instructions in the data area, comes from TRA, BCCH connection is established. There are many publications about covert channels in ICMP and IP protocols, some of which are by authors in [46, 47, & 48]. For BCCH security, transmitted data would be encrypted using best available cryptosystem. Also encryption with symmetric keys set in advance in TRM and TRA is possible. For clarity purpose, for instance, before 'packaging', when packing data into the payload of ICMP packets, quite strong encryption would be applied to the data, using, for example, 3DES, AES-256 or Blowfish, to prevent them from 'eavesdropping' or takeover by network sniffers on the transmission path. This encrypted data is then decrypted by TRA, and instruction contained therein is carried out. TRA, in turn, encrypts all responses to TRM. Furthermore, ICMP tunneling can be categorized as an encrypted communication channel between two communicating devices as without proper deep packet inspection or log review, administrators cannot detect this type of traffic through their network [49].

The case is different when data would be transferred using TCP/IP, when any applicable cryptosystem could be used to encrypt the data but this is not being considered as priority as the main purpose of the TR mechanism is to prevent third party from having copy of confidential data being transferred (encrypted or not). If the copy of confidential data (encrypted or not) is available to third party during transfer, then the main purpose of the TR mechanism itself is defeated. Preventing third party from having copy of confidential data is of utmost importance because, as earlier explained, very strong encryption nowadays may eventually become weak encryption.

Furthermore, encryption of confidential data itself applies if one cares about data cryptographic protection, otherwise 'plaintext' option, which of course means no encryption, can be applied. Confidential data is being transferred among RDt, only through a trusted route (route comprising of only RDt). In addition, route is only trusted if the two entities being connected, e.g. TRM and TRA or TRA and TRA, are trusted. By any change to the trusted channel, due to changes in any of the trusted entities, the channel is dropped and excluded by excluding any compromised RDt from among those participating in data transfer. Moreover, confidential data are being sent over channel after a trusted route is already defined, and this happens only when a route passing only through a set of RDt is defined. In addition, all RDt in trusted route are being administered already by TRM. To this, it could be said that a virtual network that is fully under the administration of TRM is generated. Above all, there exists special algorithm called probing algorithm for monitoring the status of RDt among others. This probing algorithm is described in details in section 4 when traffic routes and traffic flow control challenges are explained.

For BCCH communication protocol, all fields, including headers, must be recorded in the system of *big-endian* (for compatibility of communication modules). Packet forwarding can be determined by the type of ICMP-packet (request/reply), but for simplicity, it makes sense for types of special packets to be uniform. Types of these special packets are depicted in table 1.

For convenience, all fields can be regarded as consisting of four bytes and type of special packet can be placed in the second 4-byte data blocks, i.e. with offset 4. Thus, field with offset 0 is ‹‹magic number›› while that with offset 4 is special packet type. Offset 8 is for special packet ID. Assigning other fields depend on the type. Required types with details about other offsets are listed in table 1.

**Table 1: Special Packet Required Type**

| Designation  Instruction description | Field Offset Type | | | | |
|---|---|---|---|---|---|
| | Special packet | | Extra options | | |
| | Type | ID | 12 | 16 | 20 |
| Node verification request | A | 1 | Node ID | | |
| Response to node verification request | R | 2 | Response | | |

| Designation | Field Offset Type | | | | |
|---|---|---|---|---|---|
| | Special packet | | Extra options | | |
| Instruction description | Type | ID | 12 | 16 | 20 |
| Data read request | A | 3 | Node ID | | |
| Response to data read request | R | 4 | Data | | |
| Static routing implementation setup | A | 3 | Node ID | Initial IP-address | Destination IP-address |
| Response to static routing implementation setup | R | 4 | Response | | |
| Data request from DB | A | 5 | Data ID | | |
| Response to data request from DB | R | 6 | Data | | |
| Data storage to DB | A | 7 | Data ID | Data | |
| Response to data storage to DB | R | 8 | Data ID | | |
| Node configuration control | A | 9 | Node ID | | |
| Response to node configuration control | R | 10 | Response | | |

Note: in table 1, special packet type can either be «A» for instruction to be carried out or «R» for response to instruction.

Instructions to be carried out are normally sent by TRM while entity that receives the instructions send back responses to them in form of instructions as well. Description of instructions considering their numbering and types are given below.

**Instruction_1. Node verification request**
This instruction is sent to RD (or RDt) from TRM to identify if the node is a trusted entity or remain such. It is of special packet type 1 and offset 12 in this special packet type 1 is the IP-address of node being verified.
Instruction is being used during extended node identification stage as well during process of monitoring status of node when redirecting traffic.

**Instruction_2. Response to node verification request**
This instruction is sent to TRM from or RDt as a response to node verification request (there is no special instruction from RD when TRA is not resident in it as then the response from the RD is just usual echo reply, which TRM treats as instruction that the node is no longer or has not been a trusted entity).
Whenever RDt sends response, it places the data being sent starting from offset 12 in this special packet type 2. Instruction is being used as well during extended node identification stage and during process of monitoring status of node when redirecting traffic.

**Instruction_3. Data read request**
This instruction is sent to RD (or RDt) from TRM. In special packet type 3 by which request to read data from OS memory is made, it is necessary to restrict the size, taking into account possible size of IP packet. Offset 12 in special packet type 3 is the IP-address of node from which data is being copied/read.
Instruction is being used practically at all stages of implementing TR mechanism.

**Instruction_4. Response to data read request**
This instruction is sent to TRM from RD or RDt as a response to data read request. Whenever RDt sends response, it places the data being sent starting from offset 12 in this special packet type 2.
Instruction is being used practically at all stages of implementing TR mechanism.

**Instruction_5. Static routing implementation setup**
This instruction is sent to RDt by TRM whenever TRI has to redirect traffic by means of static routing. In this special packet type 5, offset 12 is the ID of the node on which to setup static routing, offset 16 is the initial IP-address that is also the node while offset 20 is the destination IP-address (the address to which the node has to forward traffic to). Instruction is being used during forced routing stage of implementing TR mechanism.

**Instruction_6. Response to static routing implementation setup**
This instruction is sent to TRM from RDt as a response to static routing implementation request. RDt places the data being sent as response starting from offset 12 in this special packet type 6 like other responses. When the response is not in accordance with requirements, the node is regarded as untrusted entity, and necessary measures are taken. Instruction is being used during forced routing stage of implementing TR mechanism.

**Instruction_7. Request to DB**
This instruction is sent to DM by TRM to get data. In this special packet type 7, offset 12 is the ID of the data being requested for.

Instruction is being used practically at all stages of implementing TR mechanism.

**Instruction_8. Response to request from DB**
This instruction is a response to request to DB, and it has data being requested for starting from offset 12. Instruction is being used practically at all stages of implementing TR mechanism.

**Instruction_9. Data storage to DB**
This instruction is normally sent to DB for storage of data. In this special packet type 9, data being stored is placed starting from offset 16 while the data ID itself is at offset 12.

Instruction is being used practically at all stages of implementing TR mechanism.

**Instruction_10. Response data storage to DB**
This special packet type 10 is sent as response to data storage to DB. It has data at offset 12 that indicates whether process is successful or not.
Instruction is being used practically at all stages of implementing TR mechanism.

**Instruction_11. Request to change node configuration**
This instruction is normally sent to RDt to effect some necessary changes in the interest of trusted routing, for example, changes like enabling/disabling source routing, enabling/disabling static routing, disabling/removing TRA etc. In this special packet type 11, ID of the node where changes has to be effected is placed at offset 12.

Instruction can be used practically at all stages of implementing TR mechanism, especially when RDt is compromised.

**Instruction_12. Response to request to change node configuration**

This special packet type 12 is sent as response to request to change node configuration. Like response to data storage to DB, it has data at offset 12 that indicates whether process is successful or not.

Instruction can be used practically at all stages of implementing TR mechanism, especially when RDt is compromised.

### 3.4. Trusted routing manager's operational challenges
TRM remotely controls RD via BCCH. It provides management console for operator and sends special messages via BCCH to RD. Also it retrieves and prints special messages received and maintains, stores, processes information about used RD and actions performed.

From TRM, packets to be processed should not be fragmented so as to avoid necessity to reassemble the packet in the RD. Also there must be uniformity as per packet type for compatibility of the modules as well as means to recognize special packet so that it is easily singled out by applicable module. Thus, ‹‹magic number›› (for example, 0xD0C3C0FD which can be regarded as sequence of abbreviations and interpretations from D0 check 3rust Condition 0F Device) that is written to the ICMP-packet at offset 0 (at the beginning of the data) can be used to identify a special packet. Then a field indicating the type of the special packet can be placed after this.

## 4. Trusted routes and traffic flow control challenges in TraConDa

The process of embedding modules, TRA, into RD OS implies, as well, making the RD a trusted entity, RDt, since TRM will now have *full control over* the RD. Getting trusted route is achievable when there are some RDt already in the regenerated network as a result of embedding TRA modules into RD OS (see Fig. 3).

Thus, from the results obtained from various packet route tracing and RT access methods that are stored in DB, after resolving IP-aliases among others, a trusted route is determined by finding at least a route that passes only through a set of RDt from a packet sending point to a packet receiving point (from a sender to a specified IP-address) (see Fig. 9). Making data packets passing through the RDt is necessary to have real control over the delivery of packets from the sender to the recipient by controlling and monitoring the passage of packets. Protocol stack TCP/IP provides guaranteed delivery of packets between the sender and receiver, so in order to reduce network load and increase the data transfer rate, additional control measures are not necessary.

Traffic control during process of sending information from a given sender to a specified recipient is performed as follows:

TRM sends traffic control signal to TRA (Fig. 2). If there exist a trusted route to the destination from the sender to the recipient (Fig. 9), the packet is forwarded along the trusted route to the recipient IP-address by source routing or by using static routing (Fig. 10), otherwise the processes of identifying and gain *full control over* RD in the network (finding more active routes to the destination and converting RD within those active routes to RDt) and adding the information about new RDt to the DB are started (Fig. 3).

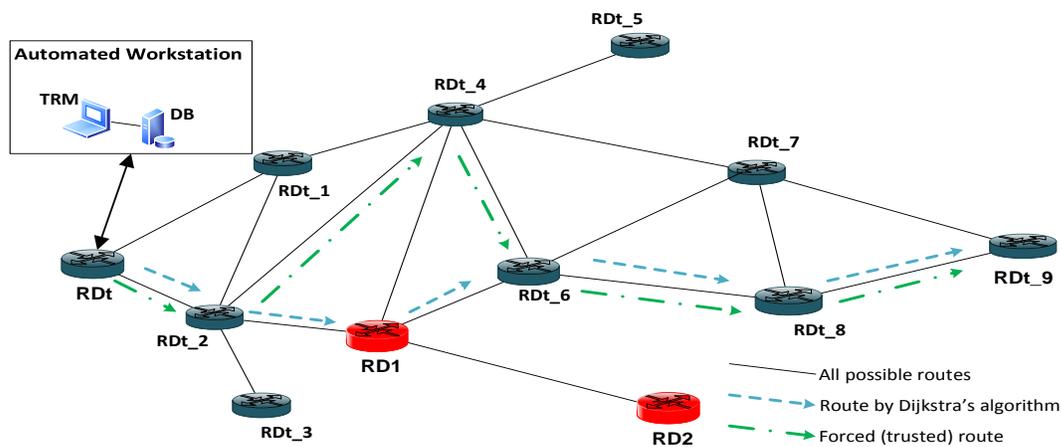

**Figure 10: Forced Routing in TraConDA**

Whenever a trusted route exists between sender and recipient, data is transferred along it by forwarding the traffic to next-hop IP-address in accordance with list of IP-addresses that comprises the trusted route. In the absence of trusted route and impossibility of getting one, message about non-availability of trusted route to the destination is sent to TRM, and this may imply starting the whole processes of TR mechanism again.

Monitoring the status of the trusted RD implies that active trusted route remains active and valid. And that the status of embedded TRA does not change during the data transfer session. Once the security of any RDt is compromised or unexpected change occurs in the RDt, the router is excluded from transferring data immediately (and data transfer is halted). Then, to continue transfer of data from the sender to the receiver, new trusted route is either recalculated or chosen from DB i.e. from among the ones found earlier during the processes of trusted route(s) calculation among all RDt within active routes (Fig. 9).

In practice, network administration concludes that a node is operational if in response to the ICMP-request came ICMP-reply from the device [50]. However, in TraConDa, during the process of data transfer, it is necessary, in addition to ICMP-request, to check periodically the existence and accuracy of the information about TRA in the OS of RD as well as general status of RDt for presence of various types of implant, critical vulnerabilities and other features that affect security of transmitted data, as well as the correctness of the active RT. Thus, the agent, TRA, guarantees that RD (OS inclusive) is "certified for TR", hence, it is trusted.

Therefore, we consider extended node identification using probe algorithm (assuming that agent is resident in node as guarantor of trustworthiness of the node).

The method is based on sending avalanche-like probes and collecting responses from them. First, TRM sends probing packets to all network nodes and each TRA processes the incoming probe by checking if it has already processed the probe earlier or not (a situation that is possible because of "looping"). If the probe was processed earlier, the TRA redistributes it to all connected node excluding only the node from which it came. Normally, TRA crosschecks the node where it is resident when a probe packet is received and generates response packet that are sent the same way as probing packet. When TRA receives the same response packet again, it stops its further redistribution. TRM collects all response packets and systemize them into identification table. Probing algorithm in TR is shown in Fig. 11.

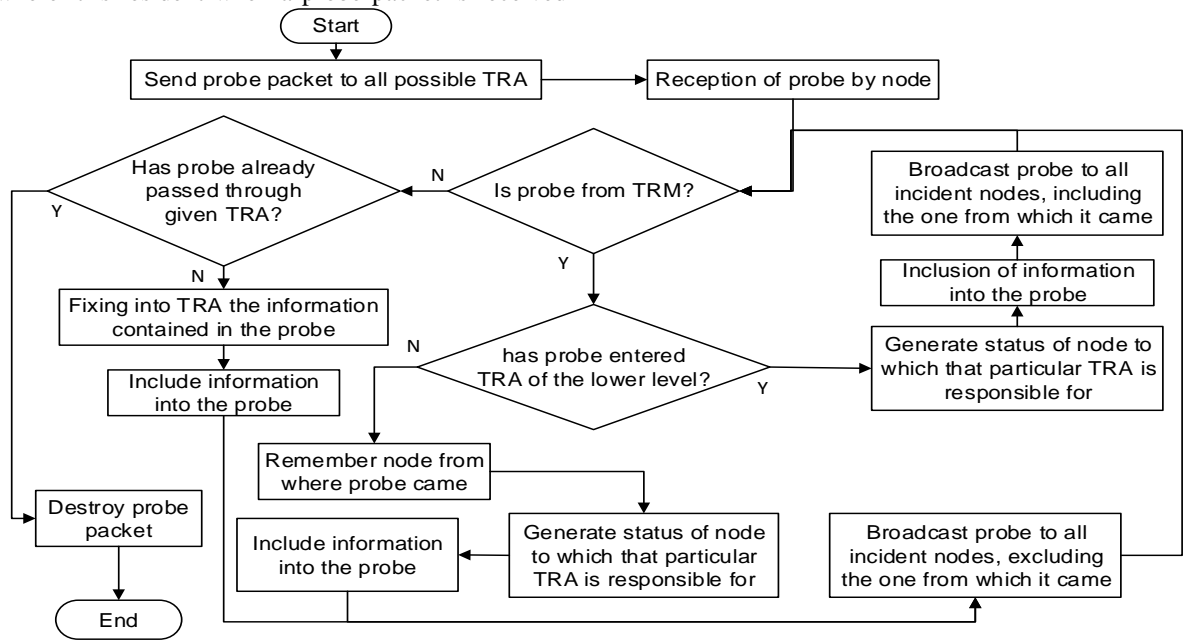

**Figure 11: Probing Algorithm in TR**

This probing algorithm allows identifying node, the existence and location of which is initially unknown, using special network packets.

## 5. Evaluation

As pure theoretical study of TraConDa is difficult as result of its extremely difficult formalization without losing the essence that is also limited only by algorithmization, and practically, requires the full implementation of the PM embedding process in GDTN (Fig. 3) and gathering of factual statistics, it is expedient to research into its properties using appropriate simulation model, a model of TN with TR. Simulation experiments were carried out and the article on imitating experiments for investigation of TR mechanism (Tiamiyu & Bunevich, 2014) indicated effectiveness of TR mechanism in TN. The simulation results showed that application of TR mechanism simultaneously improves security of TN, as well as confidentiality and identifiability of traffic. So also functional stability. Furthermore, TR mechanism simulation had shown that the more the RDt and the more the given time for data transfer, the greater the chance of the message (confidential data) reaching the destination. Considering better performance, TR mechanism simulation also had shown that for timely but secure transfer of confidential data, there is a certain optimum, when "timeliness" (i.e. given time for data transfer to the destination while increasing intensity of controlling RD (more RD to RDt)) is already quite great, and "secrecy" (i.e. confidentiality of data being transferred while increasing intensity of controlling RD) also is still great. This optimum is functional objective for TR mechanism optimization for better performance [51].

TR mechanism is also scalable like VPN and MPLS VPNs but provides for better data integrity and confidentiality. The main advantage of this proposed security mechanism, TraConDa, over other existing is that having copy of data (encrypted or not) by third party is being prevented by excluding (from participating in data transfer) compromised RD and any other RD that do not satisfy all criteria (in accordance with requirements for TR) for it to be called trusted RD.

Further, the imitating experiment aids at performing qualitative assessment of the impact of TR mechanism using SANS/GIAC method, method that evaluates "severity" of network attack that resulted from the implementation of the risk of network attack. As evaluating the effectiveness of TR mechanism in GDTN is directly related to their information security and can be, in particular, carried out through the evaluation of the risks associated with threats to network security. One such technique is generally accepted SANS/GIAC, developed at the Institute of SANS (System Administration, Networking and Security Institute) and the center of the analysis of computer incidents GIAC (Global Incident Analysis Center) [52, 53 & 54]. This technique evaluates "severity" of network attack (Sv), which is determined by the value of the risk of this attack. The magnitude of the risk is determined by the probability of the success of the attack, Lethality (Le) and the magnitude of possible damage, Criticality (Cr). The magnitude of vulnerability of TN is determined by effectiveness of system countermeasures (Sc) and network countermeasures (Nc), applied against such type of threat.
Severity of attack is determined by a numerical scale from $-10$ to 10 in accordance with the formula:
$$S_v = (C_r + L_e) - (S_c + N_c).$$
(1)

$C_r$, $L_e$, $S_c$ and $N_c$ are determined on a 5-point scale as shown in table 2.

**Table 2: Applicable SANS/GIAC 5-point scale**

| 5-point scale | $C_r$ | $L_e$ | $S_c$ | $N_c$ |
|---|---|---|---|---|

| 5-point scale | $C_r$ | $L_e$ | $S_c$ | $N_c$ |
|---|---|---|---|---|
| 5 | Data security, router, firewall, DNS server | Attack linked to obtaining root access on the remote system | modern secured operating system, installed all the software correction, the additional network protection | One method of network security, firewall, is the only entry point to the network, which implements the principle of minimizing privileges |
| 4 | Data by addressing and routing, email gateway | DoS while implementing network attack | | Several security mechanisms, firewall and presence of additional network entry points |
| 3 | Communication resources | Decline in functionality, introduction of false information, obtaining non-privileged user rights on the remote system | OS with special security features, some software correction not installed, old version of OS | Several security mechanisms with organizational measures |
| 2 | Data interactions, UNIX workstation | Disclosure of confidential information in the course of unauthorized network access | | Firewall that solves all that is not explicitly prohibited (permissive access control policy) |

| 5-point scale | $C_r$ | $L_e$ | $S_c$ | $N_c$ |
|---|---|---|---|---|
| 1 | Locally processed data, PC | Opening topological and flow structure, probability of success of attack is very small | OS without special security features, passwords are sent over the network in clear text, non-existence of passwords management policy | Only organizational measures |

**Table 3: Risk value of network attacks with and without TR**

| № | Network attack being estimated | Description | $C_r$ | $L_e$ | $S_c$ | $N_c$ | $N_{Tc}$ | $S_v$ | $S_T$ |
|---|---|---|---|---|---|---|---|---|---|
| 1. | IST-1 | Data acquisition threat by traffic analysis | 5 | 2 | 3 | 4 | 5 | 0 | -1 |
| 2. | IST-2.1 | Threat of port protocol detection | 2 | 3 | 3 | 4 | 5 | -2 | -3 |
| 3. | IST-2.2 | Threat of identification of active network services | 2 | 3 | 3 | 4 | 5 | -2 | -2 |
| 4. | IST-3.1 | Threat of substitution of trusted object of network with establishment of a virtual connection | 4 | 2 | 3 | 1 | 5 | 2 | -2 |
| 5. | IST-3.2 | Threat of substitution of trusted object of network without establishment of a virtual connection | 4 | 2 | 3 | 1 | 5 | 2 | -2 |
| 6. | IST-4.1 | Threat of imposing a false route for the purpose of intercepting traffic, for example | 4 | 2 | 3 | 4 | 5 | -1 | -2 |
| 7. | IST-4.2 | Threat of imposing a false route using, for example, ICMP protocol to disrupt communications | 2 | 4 | 3 | 4 | 3 | -1 | -2 |
| 8. | IST-5.1 | Threat of introducing false ARP-server | 5 | 3 | 3 | 3 | 5 | 2 | 2 |
| 9. | IST-5.2 | Threat of introducing of false DNS-server by intercepting DNS-request | 5 | 3 | 3 | 3 | 5 | 2 | 0 |

NETWORK COUNTERMEASURES while applying TR mechanism, $N_{Tc}$, in GDTN can be assessed on a 5-point scale as follows:
• 5: liquidate, eliminate the threat of attack
• 3: partially or indirectly weakens threat
• Equals to $N_c$: out of TR mechanism's control

Based on these scales, we find, for example, $S_v$ associated with DoS attacks resulting from typical ping-flooding.

Accordingly, $C_r = 3$, $L_e = 4$, $S_c = 3^*$, $N_c = 3$ и $N_{Tc} = 5$ (* — in TR mechanism, $S_c$ is assumed equals to 3 for all case (i.e. for all cases, average magnitude is chosen), moreover $S_c$ is outside TR mechanism's control.

Applying formula 1:
$S_v = (3 + 4) - (3 + 3) = 1$
Thus, $S_v$ in this case is 1 as there are special tools against such network attack (notwithstanding, this attack needs attention of expert).
Further, we find $S_T$ that is $S_v$ for the same attack when TR mechanism is applied
$S_T = (3 + 4) - (3 + 5) = -1$.
Severity, associated with DoS attacks resulting from typical ping-flooding declined further with the application of TR.
Similarly are calculated $S_v$ and $S_T$ in relation to other IST in GDTN. The results are shown in table 3.

| № | Network attack being estimated | Description | $C_r$ | $L_e$ | $S_c$ | $N_c$ | $N_{Tc}$ | $S_v$ | $S_T$ |
|---|---|---|---|---|---|---|---|---|---|
| 10. | IST-5.3 | Threat of introducing of false DNS-server by storming DNS-responses in computer network | 5 | 3 | 3 | 3 | 5 | 2 | 2 |
| 11. | IST-5.4 | Threat of introducing of false DNS-server by storming DNS-responses in DNS-server | 5 | 3 | 3 | 3 | 5 | 2 | 2 |
| 12. | IST-6.1 | Threat of reducing bandwidth and network performance by Ping flooding, ICMP-flooding, FTP-flooding | 3 | 4 | 3 | 3 | 5 | 1 | -1 |
| 13. | IST-6.2 | Threat of exhausting network resources by SYN-flooding and messages type Smurf and Spam | 3 | 4 | 3 | 3 | 5 | 1 | -1 |
| 14. | IST-6.3 | Threat of violating logical network connectivity by changing RT or identification/aut hentication information | 3 | 4 | 3 | 3 | 5 | 1 | -1 |

| № | Network attack being estimated | Description | $C_r$ | $L_e$ | $S_c$ | $N_c$ | $N_{Tc}$ | $S_v$ | $S_T$ |
|---|---|---|---|---|---|---|---|---|---|
| 15. | IST-6.4 | Threat of failure or malfunctioning of network services by directional storming query or transfer of non-standard packages | 3 | 4 | 3 | 3 | 5 | 1 | -1 |
| 16. | IST-7 | Threat of remote starting of applications | 1 | 3 | 3 | 1 | 3 | 0 | -4 |
| 17. | IST-8 | Threat of hardware failures in information and telecommunicati on systems | 3 | 4 | 3 | 1 | 3 | 3 | -1 |
| 18. | IST-9 | Threat of software failures in information and telecommunicati on systems | 3 | 4 | 3 | 1 | 3 | 3 | -1 |

Severity, associated with network attacks in GDTN is shown in Fig. 12.

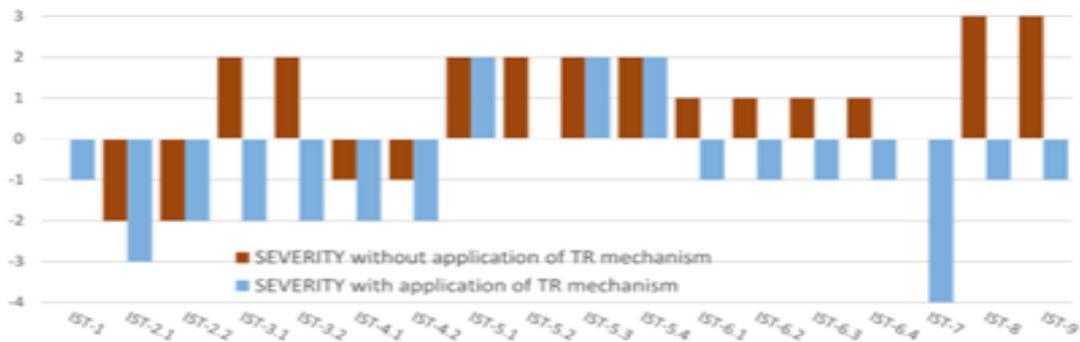

**Figure 12: Security Evaluation in GDTN with and without TR**

As seen in Fig. 12 and table 3, applying this mechanism, the vast majority of the threats are either liquidated or eliminated. Details in table 4.

**Table 4: Impacts of TR on various types of Network Threats**

| IST type | Possibility of elimination in TR mechanism |
|---|---|
| IST-1, IST-3.1, IST-3.2, IST-4.1, IST-5.1, IST-5.2, IST-5.3, IST-5.4 | Eliminated by TR by "controlling" or redirect traffic only in accordance with package trusted routes |
| IST-6.1, IST-6.2, IST-6.3, IST-6.4 | Eliminated by load control on routers and "controlling" of nearby network nodes |
| IST-4.2, IST-7, IST-8, IST-9 | Can be eased indirectly by redirecting traffic on a more sustainable/reliable route |
| IST-2.1, IST-2.2 | Out of TR action/control |

Lastly, advantages of the proposed TraConDa that is based on TR mechanism are evaluated from the view point of some of the requirements.

**Requirement 1:** Identification of trusted nodes:

Once it is known that the device is a router, and the kind of running OS, it is necessary to decide on whether it is a trusted RD, i.e. RDt or not. RD is trusted if already TRA is resident in it. Though it can be assumed that a RD is trusted based on other criteria. In the RU Patent № 150245 (2015) on trusted routing device in telecommunication networks — ustroystvo d_overennoy marshrutizatsii v telekommunikatsionnykh setyakh (Tiamiyu, 2015) are mentioned some of these criteria. But as there is necessity to have full control over RD participating in transmission of data during process of TR mechanism, for TraConDa, RD is accepted as trusted when TRA resides in it.

**Requirement 2:** Definition of trusted route:

From among RDt within active routes, a route from a source to a destination is sourced for analytically or using modified Dijkstra's algorithm [33]. Such route passing only through a set of RDt is a trusted route as traffic through such route is fully under control of TRM located at AW. Trusted routes passing through only trusted nodes (that are as well remotely controllable and under monitoring) are obtained by gaining full control on untrusted nodes in the active routes within regenerated network (when necessary and where it is possible) to prevent unauthorized access.

**Requirement 3:** Monitoring and control over trusted RD and trusted route:

TRM controls information flows via trusted routes as well as the status of trusted routes by monitoring the status of the RDt to ensure that RDt regularly transmit data according to their RT and as instructed via command from TRM, and that TRA that resides in RDt continue functioning as stipulated.

## 6. Recommendation

Considering the structure of OSPF, it is possible to modify SPF algorithm therein. Thus, modified Dijkstra's algorithm can be exchanged for SPF algorithm in OSPF in other to have modified OSPF (i.e. OSPF based on trust) that is then used to obtain trusted routing table for data transfer. As finding best path/route using OSPF protocol is actually done by finding such in topological table that is a road map of any given network topology and that contains all possible routing information paths (routes/networks), additional column can be added to topological table where information about trustworthiness of each RD in the topological table could be stored.

Of course, this entails finding extra information, information to decide the trustworthiness of RD, while collecting information about each RD on a network (i.e. during topological table build up). This extra information can be stored as Boolean values of YES or NO (True or False) to indicate existence and non-existence of criteria for the RD to be tagged as trusted entity. Thus, the existence of such column in topological table allows modified Dijkstra's algorithm to calculate trusted route. In essence, applying this, routing in the resulted network would be based on trust that is based on the trustworthiness of participating RD. However, participating RD have to meet certain trust requirements among which could be the presence of PM in the form of agent or other criteria that make the RD a trusted entity. Further studies on TraConDa could be based on this and other possible ways of implementing this TR mechanism in GDTN.

# 7. Discussion

TraConDa, based on TR mechanism, is the process of planning the transfer of information flow on the calculated route through TN nodes, excluding the possibility of tampering with the information in any form while the information stream is passing through the TN nodes [5]. Thus, TraConDa is for remote control of RD to allow for traffic redirection in the interest of TR (i.e. preventing unauthorized access to data being transferred) within GDTN, and it is structured functionally so as to guarantee and maintain data integrity and confidentiality while data passes through GDTN from a sender to the recipient. Of course, proposed TraConDa has agent, TRA, that will reside in remote RD but the functions of this agent is strictly for the purpose of ensuring data integrity and confidentiality by monitoring the status of the remote RD and its RT to provide information necessary for one of components of TraConDa, the TRM, so that it can plan traffic redirection (when necessary) of the packets originating from AW where TRM resides. The agent, TRA, will not tamper with any other packets except the ones from its owner, thus, its presence poses no threat of any kind to the owner and users of the RD. RD where TRA is resident is referred to as RD under full control of TRM and thus, it is a RDt.

Importance of development of special software complex for TN traffic control in the interest of TR cannot be over-emphasized considering recent development in security of data over networks like Internet. The set of techniques and realization methods of TR mechanism on which TraConDa is based allow:

I. For secure transfer of information from the sender to the recipient.
II. Dynamically generate active trusted routes between sender and receiver.
III. Use secure BCCH for control of data transmission.
IV. Gain full control over RD by embedding TRA in its OS, which guarantees functioning of RD in the interest of TR.

Architectural and functional design of TraConDa is for control of traffic flows to avoid tampering with data in any form, thereby guaranteeing confidentiality and integrity of the data while making the data available to the recipient. In other words, the main objective of architectural and functional design of TraConDa is to prevent third party from having copy of transmitted data.

More precisely, illegal confidential data copying is prevented in RD by avoiding or excluding compromised RD that, as a result, are not trusted (and cannot be made a trusted entity). This exclusion further prevents network attacks of various type during confidential data transfer process, among which are session hijacking, DoS, DDoS and Man-in-the-middle attacks.

In this work, it is accepted that trust relationships between nodes is established only between two RDt, as any line of connection between two trusted nodes is regarded as trusted channel. In addition, as TRM automatically becomes remote administrator of those RDt, all processes are being monitored and controlled. Sending data from a RDt to another RDt that is directly connected to it is secure. Therefore, a trusted route among many possible routes between the sender and recipient is chosen and used in routing data (source routing). This implies that data transfer is via trusted channels among RDt.

Furthermore there are always tens of other RD within GDTN, e.g. Internet, to replace 'compromised' RDt. Lastly, the status of RDt should not necessarily be permanent, i.e. keeping RD under control of TRM is a scenario that can be changed as RD can be released as soon as it is no longer needed (when confidential data already sent) by replacing the modified OS with the original (unmodified) OS of the RD.

# 8. Conclusions

Data encryption is good when considering data integrity and confidentiality but advancement in technology could render secrecy of encrypted data useless, thus the proposed TraConDa that is based on TR mechanism is very appropriate and good means to an end for those who cherish confidentiality and integrity but cannot do away with Internet when transferring information since the methods therein inhibit third party having access or copy the data being transmitted, encrypted or not. No copy of data, nothing to manipulate or decrypt now or later which implies that the data integrity and confidentiality are maintained while making the data available to the recipient. Moreover, requirements for TR mechanism on which TraConDa is based can be met easily, using standard tools and standard procedures.

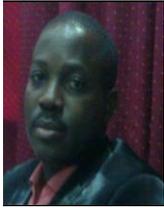

**Tiamiyu, Osuolale Abdulrahamon:** A computer network and telecommunications specialist by profession; with good multinational working experience and achievements in companies like Motorola, EXQU ltd Nigeria and several academic institutions. He holds MSc & BSc degrees in Engineering "Informatics and Computer Engineering, and presently, he is a doctoral researcher at Saint-Petersburg State University of Telecommunications, Russia. Already, he published many journal articles and his research interest is on Telecommunications Network and Network Security. Until now, he was HOD, dept. of Telecommunication Science, university of Ilorin.
Email: ozutiams@yahoo.com